\documentclass[amsmath,amssymb]{elsart3}

\usepackage{graphicx}

\usepackage{amssymb}

\begin{document}

\begin{frontmatter}



\title{Theory of RIXS in strongly correlated electron systems:\\Mott gap excitations in cuprates}


\author[label1]{T. Tohyama}
\author[label1]{K. Tsutsui}
\author[label1,label2]{S. Maekawa}

\address[label1]{Institute for Materials Research, Tohoku University, Sendai 980-8577, Japan}
\address[label2]{CREST, Japan Science and Technology Agency (JST), Kawaguchi, 
Saitama 332-0012, Japan}

\begin{abstract}
We theoretically examine the momentum dependence of resonant inelastic x-ray scattering (RIXS) spectrum for one-dimensional and two-dimensional cuprates based on the single-band Hubbard model with realistic parameter values.  The spectrum is calculated by using the numerical diagonalization technique for finite-size clusters.  We focus on excitations across the Mott gap and clarify spectral features coming from the excitations as well as the physics behind them.  Good agreement between the theoretical and existing experimental results clearly demonstrates that the RIXS is a potential tool to study the momentum-dependent charge excitations in strongly correlated electron systems.
\end{abstract}


\end{frontmatter}

\section{Introduction}
\label{Intro}

There are two-types of insulators: One is a band insulator and the other is a Mott insulator.  In the band insulator, the charge gap is controlled by the energy separation of the valence and conduction bands.  Control of the gap magnitude as well as the momentum dependence of the band dispersions is crucial for the development of semiconductor devices.  On the other hand, the charge gap in the Mott insulator is a consequence of strong electron correlation.  Two bands separated by the gap are called the lower Hubbard band (LHB) and upper Hubbard band (UHB).  The nature of the momentum dependence of the two bands will also be important for the application of the Mott insulator to future devices.

Insulating cuprates are a good example of the Mott insulator.  The dispersion of the LHB (more precisely the Zhang-Rice band) has been examined by angle-resolved photoemission spectroscopy (ARPES) and a plenty of information has been accumulated.  On the contrary, the dispersion and spectral properties of the UHB have not directly been observed so far.   In the last five years, resonant inelastic X-ray scattering (RIXS) in which the incident photon energy is tuned through Cu $K$ absorption edge has been gaining importance as a powerful technique for the investigation of the momentum-dependent Mott gap excitations in one-dimensional (1D)~\cite{Hasan1,Kim1,Suga} and two-dimensional (2D)~\cite{Hasan2,Kim2,Kim3,Hasan3,Ishii1,Ishii2,Lu} cuprates.  However, in order to obtain the information of the UHB, the experimental data themselves are not enough due to a particle-hole nature in RIXS, so that theoretical contribution and analysis are necessary.

In this paper, we theoretically examine the momentum dependence of the Cu-K edge RIXS spectrum based on the single-band Hubbard model~\cite{Tsutsui2,Tsutsui1,Tsutsui3}.  In 1D insulating cuprates, the RIXS spectrum dominantly reflects the particle-hole excitation across a direct Mott gap, for which only the charge degree of freedom contributes because of the spin-charge separation.  In 2D insulating cuprates at half filling, there appears an anisotropic momentum dependence along the $(\pi,\pi)$ and $(\pi,0)$ directions.  This is explained by the dispersion of the UHB that has
the minimum energy at $(\pi,0)$ and by a coherence effect due to antiferromagnetic (AF) order. 
Upon hole doping into the 2D insulating cuprates, the Mott gap excitation becomes broad and less momentum dependent.  This is in contrast to the electron-doped case, where the momentum dependence of the spectrum of undoped system remains. Such a contrasting behavior is originated from the difference of AF correlation in both dopings.

The rest of this paper is organized as follows.  We introduce the single-band Hubbard model and Cu-K edge RIXS process in Sec.~\ref{RIXS}.  In Sec.~\ref{1D}, the RIXS spectrum in 1D insulating cuprates is discussed.  The 2D case is given in Sec.~\ref{2D}, emphasizing the effect of AF order.  The doping effects on the RIXS spectrum are shown in Sec.~\ref{doping}, and a contrasting behavior between hole- and electron-dopings is discussed.  The summary and perspective are given in Sec.~\ref{Sum}.

\section{RIXS for Cu-K edge}
\label{RIXS}

The electronic states near the Fermi level in cuprates are known to be described by the single-band Hubbard model.  In 2D, the hopping of electron up to third neighbors are necessary while in 1D only nearest-neighborer hopping is enough to discuss the electronic states~\cite{Maekawa}.  
The Hubbard Hamiltonian up to the third neighbor hoppings for the $3d$
electron system is written as,
\begin{eqnarray}\label{ham3d}
&H&_{3d} = -t\sum_{\langle \mathbf{i},\mathbf{j} \rangle_\mathrm{1st}, \sigma}
          d_{\mathbf{i},\sigma}^\dag d_{\mathbf{j},\sigma}
        -t'\sum_{\langle \mathbf{i},\mathbf{j} \rangle_\mathrm{2nd}, \sigma}
          d_{\mathbf{i},\sigma}^\dag d_{\mathbf{j},\sigma} \nonumber\\
      &&-t''\sum_{\langle \mathbf{i},\mathbf{j} \rangle_\mathrm{3rd}, \sigma}
          d_{\mathbf{i},\sigma}^\dag d_{\mathbf{j},\sigma} + \mathrm{H.c.}
      +U\sum_\mathbf{i}
          n^d_{\mathbf{i},\uparrow}n^d_{\mathbf{i},\downarrow},
\end{eqnarray}
where $d_{\mathbf{i},\sigma}^\dag$ is the creation operator of $3d$ electron
with spin $\sigma$ at site $\mathbf{i}$,
$n_{\mathbf{i},\sigma}^d=d_{\mathbf{i},\sigma}^\dag
d_{\mathbf{i},\sigma}$, the summations
$\langle \mathbf{i},\mathbf{j} \rangle_\mathrm{1st}$,
$\langle \mathbf{i},\mathbf{j} \rangle_\mathrm{2nd}$, and
$\langle \mathbf{i},\mathbf{j} \rangle_\mathrm{3rd}$ run over first, second,
and third nearest-neighbor pairs, respectively, and the rest of the notation
is standard.

\begin{figure}
\begin{center}
\includegraphics[width=6cm]{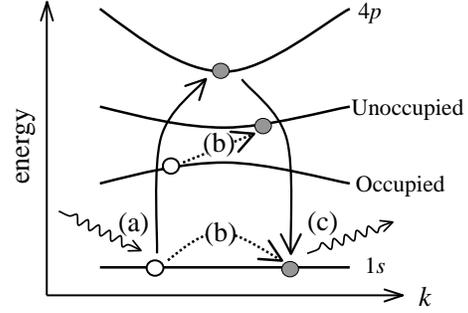}
\caption{\label{fig1}
Schematic picture of the Cu K-edge RIXS process.
An incident photon is absorbed and dipole transition $1s\rightarrow 4p$ is
 brought about [process (a)], and through the intermediate state
[process (b)], the photo-excited $4p$ electron goes to $1s$ again and
a photon is emitted [process (c)].}
\end{center}
\end{figure}

Figure~\ref{fig1} shows the schematic process of Cu $K$-edge RIXS.
An absorption of an incident photon with energy $\omega_i$, momentum
$\mathbf{K}_i$, and polarization $\mathbf{\epsilon}_i$ brings about
the dipole transition of an electron from Cu $1s$ to $4p$ orbital
[process (a)].  In the intermediate states, $3d$ electrons interact with a $1s$-core hole and
a photo-excited $4p$ electron via the Coulomb interactions so that the
excitations in the $3d$ electron system are evolved
[process (b)].  The $4p$ electron goes back to the $1s$ orbital again and a photon with
energy $\omega_f$, momentum $\mathbf{K}_f$, and polarization
${\bf\epsilon}_f$ is emitted [process (c)].  The differences of the energies and the momenta between incident and emitted
photons are transferred to the $3d$ electron system.

In the intermediate state, there are a $1s$-core hole and a $4p$ electron,
with which $3d$ electrons interact.
Since the 1$s$-core hole is localized because of a small radius of the Cu
1$s$ orbital, the attractive interaction between the 1$s$-core hole and
3$d$ electrons is very strong.
The interaction is written as,
\begin{eqnarray}\label{H1s3d}
H_{1s\mathrm{-}3d}=-V\sum_{\mathbf{i},\sigma,\sigma'}
n_{\mathbf{i},\sigma}^d n_{\mathbf{i},\sigma'}^s,
\end{eqnarray}
where $n_{\mathbf{i},\sigma}^s$ is the number operator of 1$s$-core hole
with spin $\sigma$ at site $\mathbf{i}$, and $V$ is taken to be positive.
On the contrary, since the 4$p$ electron is delocalized, the repulsive
interaction between the 4$p$ and 3$d$ electrons as well as the attractive
one between the 4$p$ electron and the 1$s$-core hole is small as compared
with the 1$s$-3$d$ interaction.
In addition, when the core-hole is screened by the 3$d$ electrons through
the strong 1$s$-3$d$ interaction,
effective charge that acts on the 4$p$ electron at the core-hole site
becomes small.
Therefore, the interactions related to the 4$p$ electron are neglected for simplicity.
Furthermore, we assume that the photo-excited 4$p$ electron enters into
the bottom of the 4$p$ band with momentum $\mathbf{k}_0$.
This assumption is justified as long as the Coulomb interactions associated
with the 4$p$ electron are neglected and the resonance condition is set to
the threshold of the 1$s$$\rightarrow$4$p$ absorption spectrum.
Under these assumptions, the RIXS spectrum is expressed as,
\begin{eqnarray}\label{rixs}
I(\Delta \mathbf{K}&,&\Delta\omega)=\sum_\alpha\left|\langle\alpha|
\sum_\sigma s_{\mathbf{k}_0-\mathbf{K}_f,\sigma} p_{\mathbf{k}_0,\sigma}
\right.\nonumber\\
&&\times\left.
\frac{1}{H-E_0-\omega_i-i\Gamma}
p_{\mathbf{k}_0,\sigma}^\dag s_{\mathbf{k}_0-\mathbf{K}_i,\sigma}^\dag
|0\rangle\right|^2
\nonumber\\&&\times
\delta(\Delta\omega-E_\alpha+E_0),
\end{eqnarray}
where $H=H_{3d}+H_{1s\mathrm{-}3d}+H_{1s,4p}$, $H_{1s,4p}$ being
kinetic and on-site energy terms  for a 1$s$-core hole and a 4$p$ electron,
$\Delta\mathbf{K}=\mathbf{K}_i-\mathbf{K}_f$,
$\Delta\omega=\omega_i-\omega_f$,
$s_{\mathbf{k},\sigma}^\dag$ ($p_{\mathbf{k},\sigma}^\dag$) is the creation operator
of the 1$s$-core hole (4$p$ electron) with momentum $\mathbf{k}$ and spin $\sigma$,
$|0\rangle$ is the ground state of the 3$d$ system with energy $E_0$,
$|\alpha\rangle$ is the final state of the RIXS process with energy $E_\alpha$, and
$\Gamma$ is the inverse of the relaxation time in the intermediate state.
In Eq.~(\ref{rixs}), the terms $H_{1s,4p}$ are replaced by $\varepsilon_{1s\mathrm{-}4p}$
which is the energy difference between the $1s$ level and the bottom of the $4p$ band.

In the resonant process, it is necessary to tune the incident photon 
energy $\omega_i$ to the energy region where Cu 1$s$ absorptions appear.
To determine such a condition, we also examine the Cu 1$s$
x-ray absorption spectroscopy (XAS) spectrum defined as,
\begin{eqnarray}\label{XAS}
D(\omega)&=&\frac{1}{\pi}\mathrm{Im}\langle 0 |
s_{\mathbf{k}_0-\mathbf{K}_i,\sigma} p_{\mathbf{k}_0,\sigma}
\frac{1}{H-E_0-\omega-i\Gamma_\mathrm{XAS}}
\nonumber\\&&\times
p_{\mathbf{k}_0,\sigma}^\dag s_{\mathbf{k}_0-\mathbf{K}_i,\sigma}^\dag
|0\rangle.
\end{eqnarray}

We calculate the RIXS and XAS spectra of Eq.~(\ref{rixs}) and (\ref{XAS}) for finite-size clusters with periodic boundary conditions by using a modified version of the conjugate-gradient method together with the Lanczos technique.

\section{One-dimensional insulating cuprate}
\label{1D}

\begin{figure}
\begin{center}
\includegraphics[width=6cm]{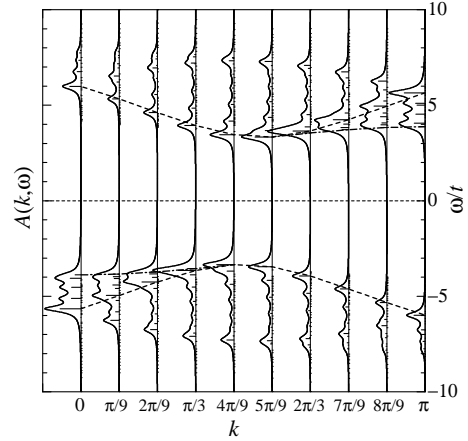}
\caption{\label{fig2}
Single-particle spectral function in the half-filled 18-site Hubbard ring with $U/t=10$.  
The $\delta$ functions are convoluted with a Lorentzian broadening of 0.2$t$.
The dotted line denotes the chemical potential.
The dot-dashed and dashed curves represent the spinon and holon (doublon) branches, respectively.}
\end{center}
\end{figure}

The dispersion of the LHB has been observed by ARPES~\cite{Kim}. The top of the band is located at the momentum $k=\pi/2$.  This is nicely reproduced by the Hubbard model with nearest-neighbor hopping.  Figure~\ref{fig2} shows the single-particle spectral function at half filling obtained by the exact-diagonalization calculation for an 18-site ring with $U/t=10$.  In the LHB below the Fermi level, we see a well-known behavior of the spin-charge separation composed of the spinon and holon branches~\cite{Kim} represented by dot-dashed and dashed curves, respectively.  The UHB also shows the two branches but the momentum region is reversed.   From the figure, we find that the Mott gap in 1D is characterized as a direct gap, indicating that a minimum charge excitation is at $\Delta K$=0.

\begin{figure}
\begin{center}
\includegraphics[width=6cm]{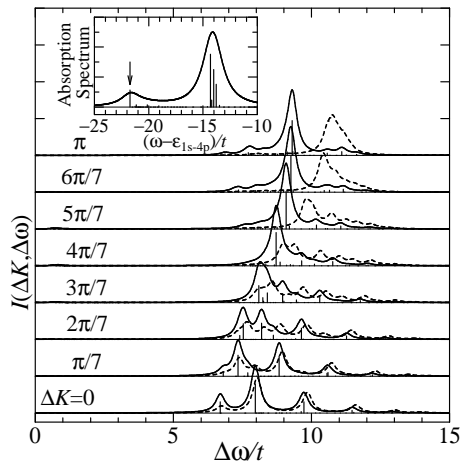}
\caption{\label{fig3}
RIXS spectra for Cu $K$-edge in a half-filled 14-site Hubbard ring.
The parameters used are $U/t$=10, $V/t$=1.5, $V_c/t$=15, and $\Gamma/t$=1.
The solid lines are obtained by the deconvolution of the $\delta$ functions with a Lorentzian broadening of 0.2$t$.
The dashed lines represent RIXS for $V/t=0$.
The inset is the Cu 1$s$ absorption spectra with a broadening of 1.0$t$, and the
 incident photon energy $\omega_i$ is set to the value denoted by the arrow.}
\end{center}
\end{figure}

The dotted curves in Fig.~\ref{fig3} exhibit RIXS spectra of a 14-site ring with $U/t=10$.
For small $\Delta K$, the spectra spread over about 8$t$.
The energy region, however, shrinks with increasing $\Delta K$, and the
 spectral weight concentrates on a narrow energy region at $\Delta K$=$\pi$.
This momentum dependence is similar to that of the dynamical
 charge-response function~\cite{Tsutsui1}, which is explained by a particle-hole model
 with only charge degree of freedom.  This is a manifestation of  the spin-charge separation~\cite{Neudert}.
It is interesting to note that the integrated weight of
 $I(\Delta K,\Delta \omega)$ with respect to $\Delta\omega$ is almost
 independent of $\Delta K$ in contrast to the charge-response function where the
 integrated weight is proportional to $\sin^2(q/2)$ for the large-$U$ region~\cite{Stephan}.

In 1D insulating cuprates, the nearest-neighbor Coulomb interaction $V$ plays a crucial role in the Mott gap excitation~\cite{Neudert}.  $I(\Delta K,\Delta \omega)$ with $V/t$=1.5 is shown in Fig.~\ref{fig3} as solid lines.  Comparing the solid lines with the dashed ones, we find that the shape of the spectrum changes remarkably for
 $\Delta K$$>$$\pi/2$, accompanied by the formation of sharp peaks
 together with the shift of the spectral weight to the lower-energy region.
This comes from exciton formation that exists at the momenta satisfying a condition $V$$>$$2t\cos (\Delta K/2)$ in the large $U$ limit~\cite{Stephan}.  
For $\Delta K$$<$$\pi/2$, the intensity of the spectrum near the
 lower edge is enhanced by the presence of excitonic effects.

The peak positions of the calculated RIXS show a dispersion consistent with that of SrCuO$_2$ observed experimentally~\cite{Hasan1,Kim1,Suga}.   
In addition to the dispersion of the peak, another dispersion with a periodicity of $\pi$ has been reported at the edge region below the exciton peak~\cite{Kim1}.  Such a feature is clearly seen in the calculated RIXS spectrum in Fig.~\ref{fig3}, although the weight is small.  We note that the edge structure does not exist unless $V$ is finite. This means that the coupling of spin and charge is recovered at finite momentum transfer in the case of finite values of $V$ and $U$.

\section{Two-dimensional insulating cuprate}
\label{2D}

\begin{figure}
\begin{center}
\includegraphics[width=6cm]{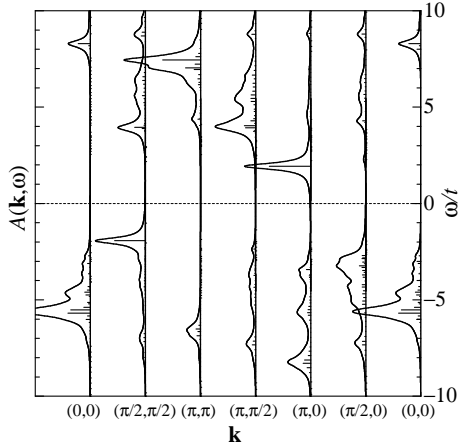}
\caption{\label{fig4}
Single-particle spectral function in a $4\times 4$ half-filled
Hubbard model with $t'$ and $t''$ terms.
$t'/t=-0.34$ and $t''/t=0.23$.
The dotted line denotes the chemical potential.
The $\delta$ functions are convoluted with a Lorentzian broadening of 0.2$t$.}
\end{center}
\end{figure}

The nature of the Mott gap excitation in 2D insulating cuprates is different from that in 1D.  Figure~\ref{fig4} shows the single-particle spectral function in a $4\times 4$ half-filled
Hubbard model with realistic values of $t'$ and $t''$.
In the LHB, the top is located at $(\pi/2,\pi/2)$, being consistent with the ARPES data for 2D insulating cuprates~\cite{Wells}.  Above the chemical potential, the dispersion of the UHB has the minimum of the
energy at $\mathbf{k}=(\pi,0)$.  Therefore, the Mott gap is indirect, in contrast to the 1D case.    
We show below that the $(\pi,0)$ spectrum in the UHB plays a crucial role in the RIXS spectrum.

\begin{figure}
\begin{center}
\includegraphics[width=6cm]{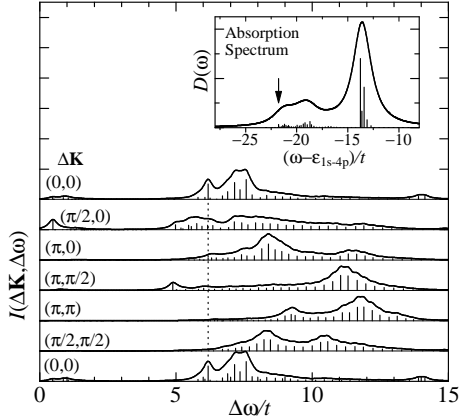}
\caption{\label{fig5}
RIXS for Cu $K$-edge in a $4\times 4$ Hubbard model with long-range hoppings.  The parameters are $U/t=10$, $V/t=15$, $\Gamma/t=1$,
$t'/t=-0.34$, and $t''/t=0.23$.
The vertical dotted line denotes the position of the peak at
$\Delta{\bf K}=(0,0)$ for guide to eyes.
The $\delta$-functions (the vertical thin solid lines) are convoluted
with a Lorentzian broadening of 0.2$t$.
Inset is the Cu 1$s$ XAS with $\Gamma_\mathrm{XAS}/t=\Gamma/t=1.0$,
and the incident photon energy $\omega_i$ is set to the value denoted by the arrow.}
\end{center}
\end{figure}

Figure~\ref{fig5} shows the momentum dependence of the RIXS spectrum.
The Mott gap excitation is above $\Delta\omega/t\sim5$. 
The spectra strongly depend on the momentum, showing a feature that the
weight shifts to higher energy region with increasing
$\left|\Delta\mathbf{K}\right|$.
The threshold of the spectrum at
$\Delta\mathbf{K}=(\pi/2,0)$ and $(\pi,\pi/2)$ is lower in energy
than that at $(0,0)$.
At $\Delta\mathbf{K}=(\pi/2,\pi/2)$, however, the spectrum appears above
the threshold at $(0,0)$, resulting in an anisotropic momentum dependence
between the spectra along $(0,0)$ to $(\pi/2,0)$ and along $(0,0)$ to
$(\pi/2,\pi/2)$.

A simple particle-hole picture based on a convoluted spectrum of the single-particle spectral function cannot explain the anisotropic momentum dependence, because the lowest-energy excitation has to appear at $\Delta\mathbf{K}=(\pi,0)-(\pi/2,-\pi/2)=(\pi/2,\pi/2)$ in the picture, being different from the calculated RIXS spectrum.  Instead of the simple argument, we consider an exact expression of the particle-hole excitation by using a two-body Green's function.  The spectral representation of the function is given by
\begin{eqnarray}\label{Bkw}
B(\mathbf{k},\Delta\mathbf{K};\omega)&=&
\sum_n\left|\langle n|\sum_\sigma
d_{\mathbf{k}+\Delta\mathbf{K},\sigma}^\dag d_{\mathbf{k},\sigma}|0\rangle\right|^2
\nonumber\\&&\times
\delta(\omega-E_\alpha+E_0),
\end{eqnarray}
where the states $|n\rangle$ have the same point-group symmetry
as that of the final states of the RIXS process.  The exact calculation of $B(\mathbf{k},\Delta\mathbf{K};\omega)$ clearly exhibits the vanishing of the $\Delta\mathbf{K}=(\pi/2,\pi/2)$ excitation~\cite{Tsutsui2}.
Note that the operator $d_{\mathbf{k}+\Delta\mathbf{K},\sigma}^\dag d_{\mathbf{k},\sigma}$ in Eq.~(\ref{Bkw}) is the same as the lowest-order contribution of $H_{1s\mathrm{-}3d}$ in Eq.~(\ref{H1s3d}). 

In order to clarify the physics behind this vanishing, we calculate $B(\mathbf{k},\Delta\mathbf{K};\omega)$ by using a standard AF spin-density- wave (SDW) mean-field approximation.   The AF-SDW mean-field Hamiltonian is given by
\begin{eqnarray}\label{HMF}
H_\mathrm{MF}=\sum_{\mathbf{k}\in\mathrm{M.Z.},\sigma}\left( E_\mathbf{k}^- \alpha_{\mathbf{k},\sigma}^\dagger \alpha_{\mathbf{k},\sigma} + E_\mathbf{k}^+ \beta_{\mathbf{k},\sigma}^\dagger \beta_{\mathbf{k},\sigma} \right) 
\end{eqnarray}
where the summation of $\mathbf{k}$ runs in the half of the original Brillouin zone enclosed by the magnetic zone (M.Z.) boundary $|k_x\pm k_y|=\pi$.  The quasiparticle dispersion is $E_\mathbf{k}^\pm=-4t'\cos k_x \cos k_y -2t''(\cos 2k_x +\cos 2k_y)\pm \varepsilon_\mathbf{k}$ with $\varepsilon_\mathbf{k}^2=4t^2(\cos k_x +\cos k_y)^2+(Um)^2$, $m$ being the staggered magnetization.  The operators $\alpha_{\mathbf{k},\sigma}$ and $\beta_{\mathbf{k},\sigma}$ are the annihilation operators of the LHB and HUB, respectively, given by 
\begin{eqnarray}\label{Bog}
d_{\mathbf{k},\sigma}&=&u_{\mathbf{k},\sigma} \beta_{\mathbf{k},\sigma} +v_{\mathbf{k},\sigma} \alpha_{\mathbf{k},\sigma} \nonumber \\
d_{\mathbf{k+Q},\sigma}&=&-v_{\mathbf{k},\sigma} \beta_{\mathbf{k},\sigma} +u_{\mathbf{k},\sigma} \alpha_{\mathbf{k},\sigma},
\end{eqnarray}
where $\mathbf{Q}=(\pi,\pi)$, $2u_{\mathbf{k},\sigma}v_{\mathbf{k},\sigma}=Um\sigma/\varepsilon_\mathbf{k}$, $u_{\mathbf{k},\sigma}^2-v_{\mathbf{k},\sigma}^2=-4t(\cos k_x +\cos k_y)/\varepsilon_\mathbf{k}$, and $u_{\mathbf{k},\sigma}+v_{\mathbf{k},\sigma}=1$.
 After some algebra, the operator in the matrix element of Eq.~(\ref{Bkw}) reads
\begin{eqnarray}\label{dd}
d_{\mathbf{k}+\Delta\mathbf{K},\sigma}^\dagger d_{\mathbf{k},\sigma}&=&\left( v_{\mathbf{k},\sigma}u_{\mathbf{k}+\Delta\mathbf{k},\sigma}-u_{\mathbf{k},\sigma}v_{\mathbf{k}+\Delta\mathbf{k},\sigma} \right) \nonumber \\
&&\times \beta_{\mathbf{k}+\Delta\mathbf{k},\sigma}^\dagger \alpha_{\mathbf{k},\sigma}.
\end{eqnarray}
The coefficient of the right-hand side is a coherence factor due to the presence of AF order.  We can easily find that, if $\mathbf{k}$ and $\mathbf{k}+\Delta\mathbf{K}$ are on the M.Z., the coherence  factor becomes zero.  This condition is satisfied for $\mathbf{k}=(\pi/2,-\pi/2)$ and  $\mathbf{k}+\Delta\mathbf{K}=(\pi,0)$.  In other words,  the particle-hole excitation from the top of the LHB to the bottom of the UHB with the momentum transfer $\Delta\mathbf{K}=(\pi/2,\pi/2)$ is forbidden due to the presence of AF order.  This is probably the reason why the RIXS spectrum at $(\pi/2,\pi/2)$ appears above the threshold at $(0,0)$ and, as a result, an anisotropic momentum dependence mentioned above emerges.  
If this is the case, the destruction of AF order by introducing hole carriers would make the RIXS spectrum more isotropic.  Actually such a behavior is seen in the doping dependence of RIXS as will be discussed in the next section.

The minimum excitation in Fig.~\ref{fig5} appears at $\Delta\mathbf{K}=(\pi/2,0)$ and $(\pi,\pi/2)$.  The excitation from the top of the LHB at $\mathbf{k}=(\pi/2,0)$ or $(0,-\pi/2)$ to the bottom of the UHB at $(\pi,0)$ contributes to the minimum as expected from the single-particle spectral function in Fig.~\ref{fig4} .  We note that  lowering the $(\pi,0)$ states in the UHB is caused by the presence of the long-range hoppings $t'$ and $t''$.  Therefore, the anisotropic feature between the $(\pi,\pi)$ and $(\pi,0)$ directions in RIXS cannot be obtained for the Hubbard model without $t'$ and $t''$.

The anisotropic behavior discussed above has been reported in RIXS experiments for Ca$_2$CuO$_2$Cl$_2$, by plotting the center of gravity of the spectral weight~\cite{Hasan2}.
In La$_2$CuO$_4$, the peak positions did not show the anisotropy~\cite{Kim2} but the onset energies of the weight did a tendency to be anisotropic~\cite{Kim3}.  The difference between the two materials would come from differences in the magnitude of the long-range hoppings that strongly influence the $\mathbf{k}=(\pi,0)$ spectral weight in the UHB. 

\section{Doping dependence in 2D}
\label{doping}

Carrier doping into the Mott insulators induces a dramatic change of the electronic states.  In high-$T_c$ cuprates, the change gives rise to asymmetric behaviors in magnetic and optical properties between hole and electron dopings~\cite{Tohyama}.  However, charge excitations across the Mott gap observed in the optical conductivity did not show such asymmetry~\cite{Uchida,Arima}.  This would be due to the fact that the optical conductivity only sees momentum-conserved excitations.  In order to clarify the momentum-dependent Mott gap excitation in doped materials, we examine the carrier and doping dependences of RIXS in this section.

\begin{figure}
\begin{center}
\includegraphics[width=6cm]{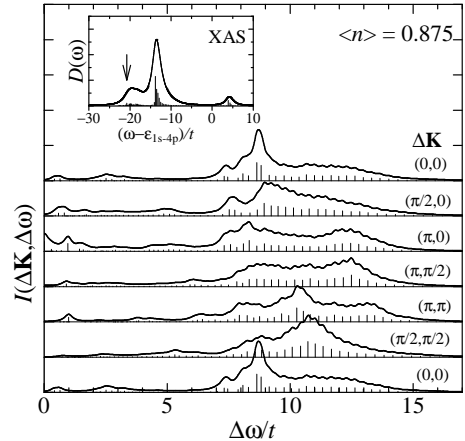}
\caption{\label{fig6}
RIXS for Cu $K$-edge in a hole-doped $4\times 4$ Hubbard model with long-range hoppings.  The electron density is $\langle n \rangle=14/16=0.875$.  The parameters are the same as those used in Fig.~\ref{fig5}.  Inset is the Cu 1$s$ XAS, and the incident photon energy $\omega_\mathrm{i}$ for RIXS is set to the value denoted by the arrow.}
\end{center}
\end{figure}

Figure~\ref{fig6} shows the RIXS spectra in the hole-doped case with the electron density $\langle n \rangle=0.875$.  
The spectra above the energy $\sim 6t$ are associated with the Mott gap excitations.
At $\Delta \mathbf{K}=(\pi,\pi)$, the spectrum extends from $\Delta\omega\sim 7t$ to $\sim 14t$.
The spectra at other $\Delta \mathbf{K}$'s are also extended to wide energy region similar to that at $(\pi,\pi)$, and thus the energy distribution of spectral weights seems to be rather independent of momentum.  Therefore, the momentum dependence is different from that at half filling shown in Fig.~\ref{fig5}.
This is consistent with the argument that the anisotropic momentum dependence of RIXS at half filling comes from AF order, because the order is rapidly destroyed upon hole doping.

\begin{figure}
\begin{center}
\includegraphics[width=6cm]{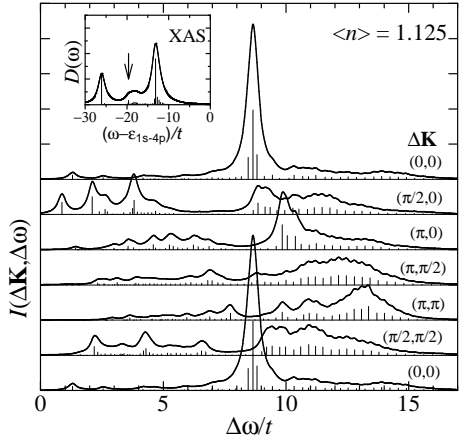}
\caption{\label{fig7}
The same as Fig.~\ref{fig6}, but electron doping with $\langle n \rangle=18/16=1.125$.}
\end{center}
\end{figure}

In contrast to hole doping, AF order remains for electron doping.  A momentum dependence similar to that at half filling is thus expected upon electron doping.  Figure~\ref{fig7} shows the RIXS spectra in the electron-doped case ($\langle n \rangle=1.125$), which are obtained by tunning the incident photon energy to the absorption peak position at around $-20t$ (see the inset).
The spectra above $8t$ are due to the Mott gap excitations.
We find that the spectra strongly depend on momentum, showing a feature that the spectral weight shifts to higher energy region with increasing $\left|\Delta\mathbf{K}\right|$.
This feature is similar to that of undoped case as discussed above.

\begin{figure}
\begin{center}
\includegraphics[width=6cm]{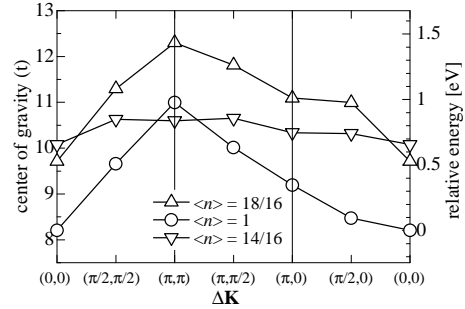}
\caption{\label{fig8}
Momentum dependence of the center of gravity of RIXS spectra associated with the Mott gap excitation.
The center of gravity is evaluated from the weight in energy region above $4t$, $6t$, and $8t$ for undoped ($\langle n\rangle=1$, from Fig.~\ref{fig5}), hole-doped (0.875, from Fig.~\ref{fig6}), and  electron-doped (1.125, form Fig.~\ref{fig7}) cases, respectively.
The right axis denotes the relative energy from the energy at $\Delta \mathbf{K}=(0,0)$ with $t=0.35$ eV.}
\end{center}
\end{figure}

In order to compare the energy positions of the spectra, the momentum dependence of the center of gravity of the RIXS spectra associated with the Mott gap excitation is plotted in Fig.~\ref{fig8}, where the spectral weight is adopted from the energy regions above $4t$, $6t$, and $8t$ for undoped ($\langle n\rangle=1$, denoted by circles), hole-doped (0.875, downward triangles), and electron-doped (1.125, upward triangles) cases, respectively.
We find that the energy positions in the doped cases are shifted to the high-energy side compared with the undoped case.
This is because, upon hole (electron) doping, the Fermi energy shifts to LHB (UHB) and also the energy difference between the occupied LHB and unoccupied UHB becomes large due to the reconstruction of the electronic states.
In the undoped case, the center of gravity shifts to higher energy with increasing $\left|\Delta\mathbf{K}\right|$.  The hole doping reduces the momentum dependence.  On the other hand, the momentum dependence in the undoped case remains by electron doping, except along the $(\pi,0)$ direction where the energy difference between $(\pi/2,0)$ and $(0,0)$ is larger than that in the undoped case.  We note that, since the $(\pi,0)$ state of the UHB is occupied by electrons upon electron doping, the center of gravity at $\Delta \mathbf{K}=(\pi/2,0)$ in the RIXS shifts to the higher energy.  

Small momentum dependence of the Mott gap excitations as compared with that at half filling has been reported for hole-doped La$_{2-x}$Sr$_x$CuO$_4$~\cite{Kim3,Hasan3} and YBa$_2$Cu$_3$O$_{7-\delta}$~\cite{Ishii1}, being consistent with the theoretical prediction.  For an electron-doped material Nd$_{0.185}$Ce$_{0.15}$CuO$_4$, recent RIXS experiments show that the Mott gap excitation concentrates on a energy ($\sim 2$~eV) at the zone center and becomes broad in energy with increasing $\Delta\mathbf{k}$.  This behavior is qualitatively similar to our theoretical results.

Finally we comment on the dependence of RIXS spectrum on the incident photon energy.  In the inset of Fig.~\ref{fig7} for electron doping, there are three absorption peaks at around $\omega-\varepsilon_{1s\mathrm{-}4p}=-26t$, $-20t$, and $-13t$.
The $-20t$ and $-13t$ peaks are also seen in the undoped and hole-doped cases (see Figs.~\ref{fig5} and~\ref{fig6}).
The peak at around $-20t$ corresponds to a final state where the core hole is screened and thus the core-hole site is doubly occupied by $3d$ electrons ($U-2U_\mathrm{c}=-20t$).  
Since this final state promotes the excitations from LHB to UHB in the RIXS, the incident photon is tuned to $\sim -20t$ to examine the Mott gap excitation.
The peak at around $-13t$ represents unscreened core-hole state and mainly contains the configuration that a core hole is created at a singly occupied site ($-U_\mathrm{c}=-15t$).
The peak at around $-26t$, which appears upon electron doping, corresponds to a final state where the core-hole is created at a doubly occupied site induced by electron doping ($-2U_\mathrm{c}=-30t$).  This means that, if we shift the incident photon energy to this peak, we can enhance charge excitations within the UHB that are seen below $\Delta \omega \sim 8t$  in Fig.~\ref{fig7}.  In fact, a RIXS experiment for Nd$_{0.185}$Ce$_{0.15}$CuO$_4$ has clearly shown such intraband charge excitations that are consistent with dynamical charge response function in the electron-doped Hubbard model~\cite{Ishii2}.  In hole doping, the absorption peak that enhances intraband excitations in RIXS is located at around $4t$.  However, tuning the incident photon energy to this region seems to be experimentally difficult because of the overlap of other absorption processes.

\section{Summary}
\label{Sum}
We have examined the RIXS for 1D and 2D cuprates based on the single-band Hubbard model with realistic parameter values.  The spectra were calculated by using the numerical diagonalization technique for finite-size clusters.  We have focused on the momentum dependence of charge excitations across the Mott gap and clarified their spectral features as well as the physics behind them.  Remarkable agreement between the theoretical and existing experimental data clearly demonstrates that RIXS is a powerful tool to study the momentum-dependent gap excitations.  
    
In order to make RIXS  a more promising tool, we would like to comment on the following two points.   1) The calculated spectral weight shown in Figs.~\ref{fig5},~\ref{fig6}, and ~\ref{fig7} are widely spread in energy above the Mott gap, but they also exhibit several broad- peak features.  These broad peaks have not been identified in the experiments.  Therefore, detailed studies on the incident photon energy dependence~\cite{Lu} as well as the improvement of energy resolutions would be important to detect these peaks. 2) In high-$T_c$ cuprates, charge inhomogeneity is considered to be important for the mechanism of superconductivity.  The role of the electron-phonon interaction has also been reexamined recently.  In the situation, the clarification of the charge dynamics in the metallic phase is crucially important.  Recent observation of the intraband charge excitation by RIXS~\cite{Ishii2} is encouraging for this direction.  However, the energy resolution ($\sim 400$~meV) is still not high enough to see the charge dynamics contributing to the superconductivity ($<100$~meV).  We therefore hope that the improvement of the resolution will  be accomplished in the near future.

\section*{Acknowledgements}
We would like to thank M. Z. Hasan, Y. J. Kim, J. Hill, L. Lu, M. Greven, K. Ishii, Y. Endoh, and S. Suga for useful discussions.  This work was supported by NAREGI Nanoscience Project, CREST, and Grant-in-Aid for Scientific Research from the Ministry of Education, Culture, Sports, Science and Technology of Japan.  The numerical calculations were partly performed in the supercomputing facilities in ISSP, University of Tokyo and IMR, Tohoku University.


\begin{thebibliography}{00}

\bibitem{Hasan1}M. Z. Hasan, P. A. Montano, E. D. Isaacs, Z.-X. Shen, H. Eisaki, S. K. Sinha, Z. Islam, N. Motoyama, and S. Uchida, Phys. Rev. Lett. {\bf 88}, 177403 (2002).
\bibitem{Kim1} Y. J. Kim, J. P. Hill, H. Benthinen, F. H. L. Essler, E. Jeckelmann, H. S. Choi, T. W. Noh, N. Motoyama, K. M. Kojima, S. Uchida, D. Casa, and T. Gog, Phys. Rev. Lett. {\bf 92}, 137402 (2004).
\bibitem{Suga}A. Higashiya, A. Shigemoto, S. Kasai, S. Imada, S. Suga, M. Sing, C. Kim, M. Yabashi, K. Tamasaku, and T. Ishikawa, Colid Syaye Commun. {\bf 130}, 7 (2004); S. Suga, S. Imada, A. Higashiya, A. Shigemoto, S. Kasai, M. Sing, H. Fujiwara, A. Sekiyama, A. Yamasaki, C. Kim, T. Nomura, J. Igarashi, M. Yabashi, and T. Ishikawa, Phys. Rev. B {\bf 72}, 081101 (2005).
\bibitem{Hasan2}M. Z. Hasan, E. D. Isaacs, Z. X. Shen, L. L. Miller, K. Tsutsui, T. Tohyama, and S. Maekawa, Science {\bf 288}, 1811 (2000).
\bibitem{Kim2}Y. J. Kim, J. P. Hill, C. A. Burns, S. Wakimoto, R. J. Birgeneau, D. Casa, T. Gog, and C. T. Venkataraman, Phys. Rev. Lett. {\bf 89}, 177003 (2002).
\bibitem{Kim3}Y. J. Kim, J. P. Hill, S. Komiya,Y. Ando, D. Casa, T. Gog, and C. T. Venkataraman, Phys. Rev. B {\bf 70}, 094524 (2004).
\bibitem{Hasan3}M. Z. Hasan, Y. Li, D. Qian, Y.-D. Chuang, H. Eisaki, S. Uchida, Y. Kaga, T. Sasagawa, and H. Takagi, cond-mat/0406654.
\bibitem{Ishii1}K. Ishii, K. Tsutsui, Y. Endoh, T. Tohyama, K. Kuzushita, T. Inami, K. Ohwada, S. Maekawa, T. Masui, S. Tajima, Y. Murakami, and J. Mizuki, Phys. Rev. Lett. {\bf 94}, 187002 (2005).
\bibitem{Ishii2}K. Ishii, K. Tsutsui, Y. Endoh, T. Tohyama, S. Maekawa, M. Hoesch, K. Kuzushita, M. Tsubota, T. Inami, J. Mizuki, Y. Murakami, and K. Yamada, Phys. Rev. Lett. {\bf 94}, 207003 (2005).
\bibitem{Lu} L. Lu, G. Chabot-Couture, X. Zhao, J. N. Hancock, N. Kaneko, O. P. Vajk, G. Yu, S. Grenier, Y. J. Kim, D. Casa, T. Gog, and M. Greven, Phys. Rev. Lett. {\bf 95}, 217003 (2005).
\bibitem{Tsutsui2}K. Tsutsui, T. Tohyama, and S. Maekawa, Phys. Rev. Lett. {\bf 83}, 3705 (1999).
\bibitem{Tsutsui1}K. Tsutsui, T. Tohyama, and S. Maekawa, Phys. Rev. B {\bf 61}, 7180 (2000).
\bibitem{Tsutsui3}K. Tsutsui, T. Tohyama, and S. Maekawa, Phys. Rev. Lett. {\bf 91}, 117001 (2003).
\bibitem{Maekawa}See, for example,  S. Maekawa, T. Tohyama, S. E. Barnes, S. Ishihara, W. Koshibae, and G. Khaliullin: {\it Physics of Transition Metal Oxides} (Springer-Verlag Berlin Heidelberg 2004).
\bibitem{Kim}    C. Kim, A. Y. Matsuura, Z.-X. Shen, N. Motoyama, H. Eisaki,
                 S. Uchida, T. Tohyama, and S. Maekawa,
                 Phys. Rev. Lett. {\bf 77}, 4054 (1996).
\bibitem{Neudert} R. Neudert, M. Knupfer, M. S. Golden, J. Fink,
                  W. Stephan, K. Penc, N. Motoyama, H. Eisaki, and S. Uchida,
                  Phys. Rev. Lett. {\bf 81}, 657 (1998).
\bibitem{Stephan}  W. Stephan and K. Penc,
                   Phys. Rev. B {\bf 54}, R17269 (1996).
\bibitem{Wells}B. O. Wells, Z. -X. Shen, A. Matsuura, D. M. King, M. A. Kastner, M. Greven, and R. J. Birgeneau, Phys. Rev. Lett. {\bf 74}, 964 (1995).
\bibitem{Tohyama}T. Tohyama, Phys. Rev. B {\bf 70}, 174517 (2004).
\bibitem{Uchida}S. Uchida, T. Ido, H. Takagi, T. Arima, Y. Tokura, and S. Tajima,
Phys. Rev. B {\bf  43}, 7942 (1991).
\bibitem{Arima}T. Arima, Y. Tokura, and S. Uchida, Phys. Rev. B {\bf 48}, 6597 (1993).

\end{thebibliography}
\end{document}